\documentclass[a4paper,10pt,reqno]{amsart}

\usepackage{amsmath, amssymb}
\usepackage{graphicx}
\usepackage{multicol}

\makeatletter

\newenvironment{figurehere}
  {\def\@captype{figure}}
  {}
\makeatother

\newtheorem{theorem}{Theorem}

\newcommand{\RE}{\mathbb{R}}
\newcommand{\erre}{\mathbb{R}}
\newcommand{\GG}{\mathcal{G}}

\newcommand{\beq}{\begin{equation}}
\newcommand{\eeq}{\end{equation}}
\newcommand{\ve}{\varepsilon}
\newcommand{\al}{\alpha}
\newcommand{\lf}{\left}
\newcommand{\ri}{\right}

\newcommand{\ome}{\omega}

\newcommand{\f}{\frac}
\DeclareMathOperator{\sech}{sech}

\DeclareMathOperator{\upd}{d}

\title{Stationary States of NLS on Star Graphs}

\author[]{Riccardo Adami}
\address{Dipartimento di Scienze Matematiche,  Politecnico di Torino,  C.so Duca degli Abruzzi 24, 
 10129 Milano, Italy, EU}
\email{ riccardo.adami@polito.it}

\author[]{Claudio Cacciapuoti}
\address{Hausdorff Center for Mathematics,
 Institut f\"ur Angewandte Mathematik, Endenicher Allee 60, 53115 Bonn, Germany, EU}
\email{cacciapuoti@him.uni-bonn.de}

\author[]{Domenico Finco}
\address{Facolt\`a di Ingegneria, Universit\`a Telematica
Internazionale Uninettuno,  Corso Vittorio Emanuele II 39, 00186 Roma,
Italy, EU}
\email{d.finco@uninettunouniversity.net}

\author[]{Diego Noja}
\address{Dipartimento di Matematica e Applicazioni, Universit\`a
 di Milano Bicocca,  via R. Cozzi 53, 20125 Milano, Italy, EU}
\email{diego.noja@unimib.it}

\begin{document}

%%%%%%%%%%%%%%%%%%%%%%%%%%%%%%%%%
%ABSTRACT
%%%%%%%%%%%%%%%%%%%%%%%%%%%%%%%%%
\begin{abstract}
We consider a generalized nonlinear Schr\"odinger equation (NLS) with a power
nonlinearity $|\psi|^{2\mu}\psi$ of focusing type describing propagation on the ramified
structure given by $N$ edges connected at a vertex (a star graph). To model the interaction at the junction, it is there imposed a boundary
condition analogous to the $\delta$ potential of strength $\alpha$ on
the line, including as a special case ($\alpha=0$) the free
propagation.  We show that nonlinear stationary states describing solitons sitting at the vertex exist both for attractive ($\alpha<0$, representing a potential well) and
repulsive ($\alpha>0$, a potential barrier) interaction. In the case of sufficiently strong attractive
interaction at the vertex and power nonlinearity $\mu<2$, including the standard cubic case, we characterize the
ground state as minimizer of a constrained action and we
discuss its orbital stability. Finally we show that in the free case,
for even $N$ only, the stationary states can be used to construct
traveling waves on the graph.

\vspace{10pt}

\noindent
\begin{footnotesize}
PACS numbers:  05.45.Yv, 03.75.Kk, 42.65.Tg 
\end{footnotesize}
\end{abstract}

\maketitle

\begin{multicols}{2}

\section{Introduction}
The Nonlinear Schr\"odinger Equation (NLS) is a well known model for
several physical systems, and a paradigm for the behaviour of
nonlinear dispersive equations \cite{SS}.  We recall two main fields
of use of NLS in physics and its applications, to which we refer in
the following: propagation of electromagnetic pulses in nonlinear
media (typically laser beams in Kerr media or signal propagation in
optical fibers), and dynamics of Bose-Einstein condensates (BEC). It
is well known that under some conditions (e.g. power nonlinearities)
NLS admits symmetries, which in turn are associated to the existence
of so called solitary solutions, usually in the form of standing or
travelling waves. Symmetries are heavily affected and usually broken
by the presence of inhomogeneities of various type, but some of them
persist and give rise to quite interesting phenomena, such as defect
induced modes \cite{CM,F}, i.e. standing solutions strongly localized
around the defect. For example the presence of defect modes affects
propagation allowing trapping of wavepackets, as experimentally shown
in the case of local photonic potentials in \cite{Linzon}; and on the
other hand, nonlinearity can induce escaping of solitons from
confining potentials, as demonstrated in \cite{Pe}, or enhanced
quantum reflection of matter wave solitons \cite{LB}. A last
interesting phenomenon is the strong alteration of tunneling through
potential barriers in the presence of nonlinear defocusing optical
media \cite {Wan}. In this Letter we consider NLS propagation through
junctions in networks showing analytically that defect or better
``junction" modes arise, and giving their explicit construction and
main properties. When the dynamics of a BEC takes place in
essentially one dimensional substrates (``cigar shaped'' condensates) or
a laser pulse propagates in optical fibers and thin waveguides, the
question arises as to  the effect of a ramified junction on propagation
and on the possible generation of stable bound states. The study of the
behavior of NLS on networks is a not yet developed but growing
subject. Concerning situations of direct physical interest we mention
the analysis of scattering at Y junctions (``beam splitters'') and other
network configurations (``ring interferometers'') for one dimensional
Bose liquids discussed in \cite{TOD}. Interplay between propagation in
photonic T-junction and nonlinearity is studied in \cite{BS}. Some more results are known for the discrete
chain NLS model (DNLS), see in connection with the present paper
\cite{Miro}. Other recent developments are  in \cite{S} and in \cite{GSD}, where in particular 
stationary scattering from complex networks sustaining nonlinear
Schr\"odinger propagation is studied.   Again regarding scattering, we mention that the behaviour of fast solitons colliding with a Y-junction was studied in \cite{ACFN}. Here we would like to give a first rigorous determination of defect modes at
junctions, i.e. nonlinear bound states, fully taking in account nonlinearity of the model NLS
equation, and some properties of the corresponding standing waves, in
particular their stability. Even if the mathematical model of a
ramified structure here used is oversimplified compared to a realistic
experimental setup (but it is the common one e.g. in the theoretical
description of quantum wires), it allows to grasp the main effect we
want to exhibit, i.e. the birth of genuinely nonlinear bound states.
We recall that the {\it linear} Schr\"odinger equation on graphs has been for a long time a very developed subject \cite{EKK}, due to its
applications in quantum chemistry, nanotechnologies and more generally
mesoscopic physics.  We consider the simple configuration of a {\it
  star graph} $\GG$ made of $N$ half lines with common origin. The
associated Hilbert space for the  Schr\"odinger dynamics is
$L^2(\GG)=\bigoplus_{j=1}^NL^2(\RE^+)$.  Its elements will be
represented as vectors functions with square integrable components,
namely
\beq
\Psi=\left(\psi_1,\dots, \psi_N\right)^T
\eeq

To define a dynamics on the graph, we split it into a linear part described by a Hamiltonian operator and a nonlinear part. To assign the Hamiltonian operator on the graph one gives it on every edge; moreover a boundary condition at the 
vertex will be needed to assure a unitary dynamics. As an operator, the Hamiltonian will be correctly defined through both the action on every edge and the boundary condition at the vertex. So it will embody the parameters possibly present in the boundary condition. To be specific, we assume that the linear part of the Hamiltonian is the selfadjoint operator $H_{\al}$ (see e.g. \cite{EKK}) which acts as the free Laplacian on every edge, $\left(H_{\al} \Psi\right)_i =-\psi_i''\ ,$ and with the following boundary conditions containing the parameter $\alpha\in \RE$
\beq
\label{domdelta}
\psi_1 (0) =  \ldots =\psi_N (0), \quad \sum_{k=1}^N \psi_k '= \alpha \psi_1 (0)\ . 
\eeq

\noindent
$H_\alpha$ generalizes to the graph the well known Schr\"odinger
operator with delta potential of strength $\alpha$ on the line, and similarly to that case, the interaction is
encoded in the boundary condition. For example the here mostly interesting case $\alpha<0$ can be interpreted as the presence of a deep attractive potential well or attractive defect at the vertex, while $\alpha>0$ represents a model of potential barrier at the junction. The case $\al=0$ in \eqref{domdelta}
plays a distinguished role and defines what is usually given the name
of free (on the line one has continuity of wavefunction and its derivative and interaction disappears) or Kirchhoff boundary
condition.  More general boundary conditions can be considered to take into account properties of different physical models.
To pose the NLS on the graph we define products
componentwise, 

\beq
\label{nonlin}
\Psi\Phi\equiv
\begin{pmatrix}
\psi_1\phi_1 \\ \vdots \\ \psi_N\phi_N
\end{pmatrix}
, \quad\text{and} \quad
|\Psi|^{2\mu}\Psi\equiv
\begin{pmatrix}
|\psi_1|^{2\mu}\psi_1 \\ \vdots \\ |\psi_N|^{2\mu}\psi_N
\end{pmatrix}.
\eeq
Now it is well defined the NLS equation on the graph, 
\beq
\label{diffform}
 i \frac{d}{dt}\Psi (t) \ = \ H_{\al} \Psi (t) - | \Psi (t) |^{2\mu} \Psi (t)\,
\eeq
where $\mu >0$. This amounts to a system of $N$ scalar NLS
equations on the halfline, coupled through the boundary condition at
the origin included in the domain \eqref{domdelta}. Notice that we consider a model with generic power nonlinearity; in most of physical applications the cubic ($\mu=1$) case is studied, but other nonlinearities appear in the literature, in particular in the analysis of wave collapse in plasma physics and nonlinear optics when high nonlinearities (e.g. non Kerr) are present (see \cite{B}). Concerning general properties of the equation, for $\mu>0$  existence and
uniqueness for initial data in the relevant state spaces holds (\cite{ACFN, ACFN3}), and if $0<\mu<2$ solution exists for all times and no collapse occurs.  Finally, as in the standard NLS on the line, mass $M(\Psi)=||\Psi||^2_2$ and energy  $E[\Psi]$ are conserved, where 
\beq E[\Psi]=
\sum_{i=1}^N\bigg[\int_0^{+\infty}\frac{|\psi_i'|^2}{2}
  -\frac{|\psi_i|^{2\mu+2}}{2\mu+2}\upd x
  \bigg]+\frac{\alpha}{2}|\psi_1(0)|^2
\label{energy} \eeq
\par\noindent 
\section{Stationary states}
We
are interested in the stationary solutions of the previous equation,
by which we mean solutions of \eqref{diffform} of the form \beq
\label{stat-sol}
\Psi (t,x)=e^{i\omega t}\ \Psi_{\omega}(x)\ .
\eeq 
The parameter $\omega$ is interpreted as propagation constant in nonlinear optics and  as chemical potential in BEC.
The amplitude $\Psi_{\omega}$ satisfies
\beq
\label{stat-eq}
H_{\al}\Psi_{\omega} - |\Psi_{\omega}|^{2\mu} \Psi_{\omega} = -\ome \Psi_{\omega} \,\qquad \ome>0.
\eeq
Locally $H_{\al}$ acts as the Laplacian and so on every edge we
must seek $L^2(\RE^+)$ solutions of the equation
\beq
-\phi'' - |\phi|^{2\mu} \phi = -\ome \phi \,\qquad \ome>0.
\eeq
The most general one is given by the function
\beq \phi (\sigma, a; x) = \sigma \lf[ (\mu+1)\ome\ri]^{\f{1}{2\mu}}
\sech^{\f{1}{\mu}} (\mu \sqrt{\ome} (x-a))
\label{galestro}
\eeq
where $|\sigma|=1$ and $a\in\RE$. 
 Therefore the components $(\Psi_{\omega})_i$
of a stationary state $\Psi_{\omega}$ must satisfy
\beq
\lf(\Psi_{\omega}\ri)_i(x)=\phi(\sigma_i,a_i;x)\,.
\eeq
Now we impose boundary conditions \eqref{domdelta}.  The
continuity condition in \eqref{domdelta} immediately implies
$\sigma_1=\ldots=\sigma_N$ and $a_i=\ve_i a$ with $\ve_i=\pm1$ and
$a>0$.  
Due to the bell shape of the function $\phi$, we shall say that in the
i-th edge: there is a {\em bump} if $a_i>0$, that is, if $\ve_i=+1$;
there is a {\em tail} if $a_i<0$, that is, if $\ve_i=-1$.
Now we determine $\ve_i$ and $a$. The second boundary condition in
\eqref{domdelta} implies \beq 
\tanh (\mu \sqrt{\ome} a)\sum_{i=1}^N
\ve_i =\f{\al}{\sqrt{\ome}}\ .
\label{eq-a}
\eeq
Eq. \eqref{eq-a} gives as a first constraint that $\sum_{i=1}^N\ve_i$ must have the same sign of $\al$. That is for $\al>0$ the stationary
state must have strictly more bumps than tails while for $\al<0$ we must have more tails than bumps. For every such a configuration
condition \eqref{eq-a} fixes uniquely $a$. We choose to index the stationary states by the number $j$ of bumps.
So $\Psi_{\omega}^j$ is given by
\beq
\lf(\Psi_{\omega}^j\ri)_i(x) = 
\begin{cases}
\phi(a^j;x) & i=1,\ldots j \\
\phi(-a^j;x) & i=j+1, \ldots N
\end{cases} \label{states1}
\eeq 
\beq 
a^j = \f{1}{\mu \sqrt{\ome}} \,\text{arctanh}
\lf(\f{\al}{(2j-N)\sqrt{\ome}} \ri) \ .\label{states2}
\eeq 
In Eq. \eqref{states1} we 
 set $\sigma_i = 1$ and omitted it in the argument of $\phi$ to simplify notation. To summarize, solutions of \eqref{stat-eq} for $\al>0$
are given by $\Psi_{\omega}^j$ with $j=[N/2 + 1], \ldots , N$  and for $\al<0$ by $\Psi_{\omega}^j$ with $j=0 ,\ldots ,[(N-1)/2]$ where $[s]$ is the integer part of $s$. Stationary states for the three-edge star graph are portrayed in Fig. \ref{fig1}, for the attractive case, and in Fig. \ref{fig2}, for the repulsive case. Notice that from \eqref{states2} follows the lower bound on existence of bound states $\frac{{\al}^2}{N^2}<\omega\ .$ We can  interpret this fact for $\alpha <0$ noticing that \eqref{stat-eq} neglecting nonlinearity is the eigenvalue equation for the linear part of the hamiltonian corresponding to energy $E=-\omega$, and taking in account the known fact that the linear graph hamiltonian $H_\alpha$ has the ground state energy $-\frac{\alpha^2}{N^2}$; so the lower bound means that the nonlinear standing waves bifurcate from the vanishing wavefunction at the ground state energy of the linear problem. It can be rigorously proved that there are no solutions at all for $\frac{{\al}^2}{N^2}\geq\omega\ .$

\begin{figurehere}
\begin{center}
\includegraphics[width=0.9\columnwidth]{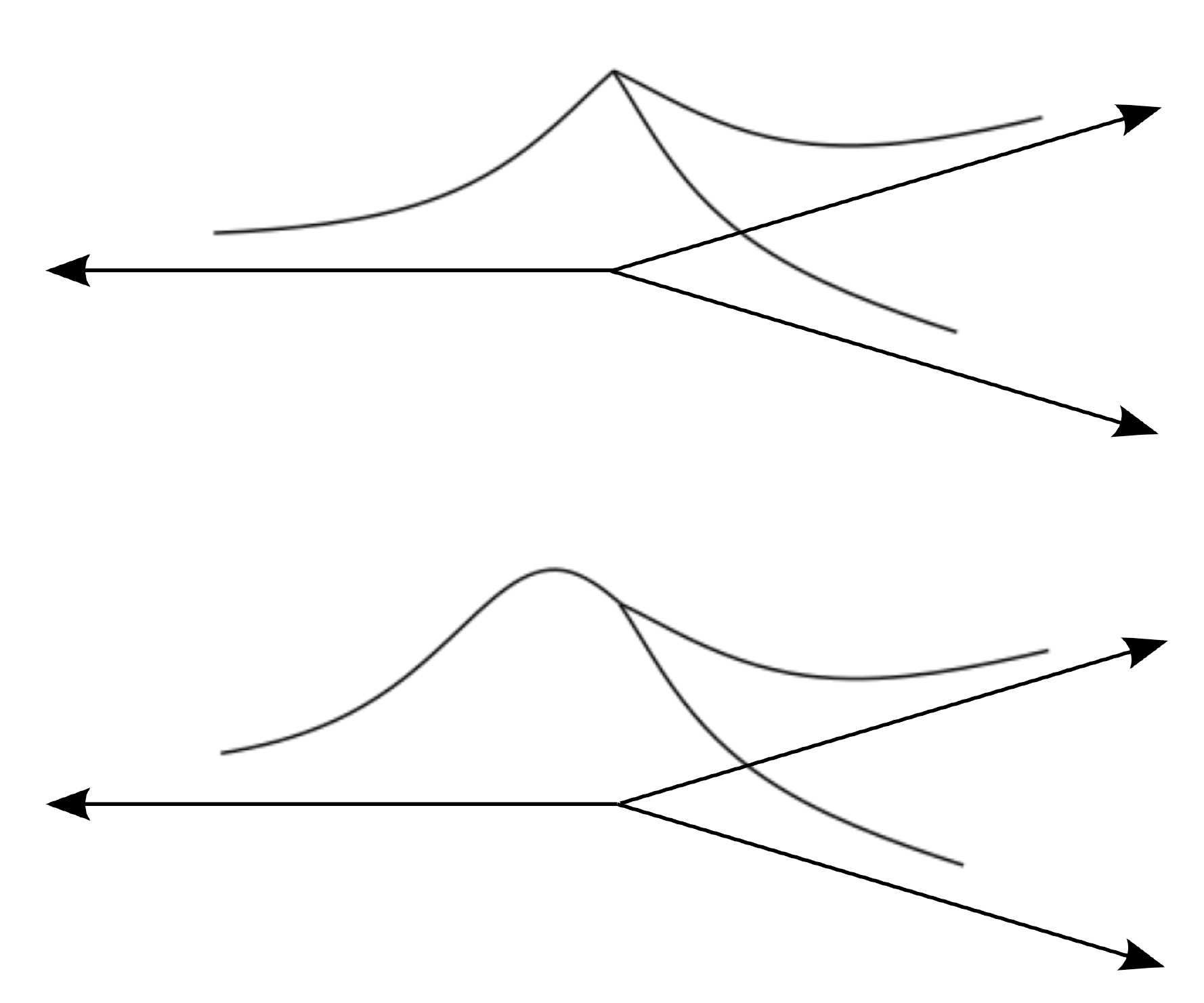}
\caption{\label{fig1}Nonlinear stationary states:  $\alpha<0, N=3, j=0,1$}
\end{center}
\end{figurehere}

\begin{figurehere}
\begin{center}
\includegraphics[width=0.9\columnwidth]{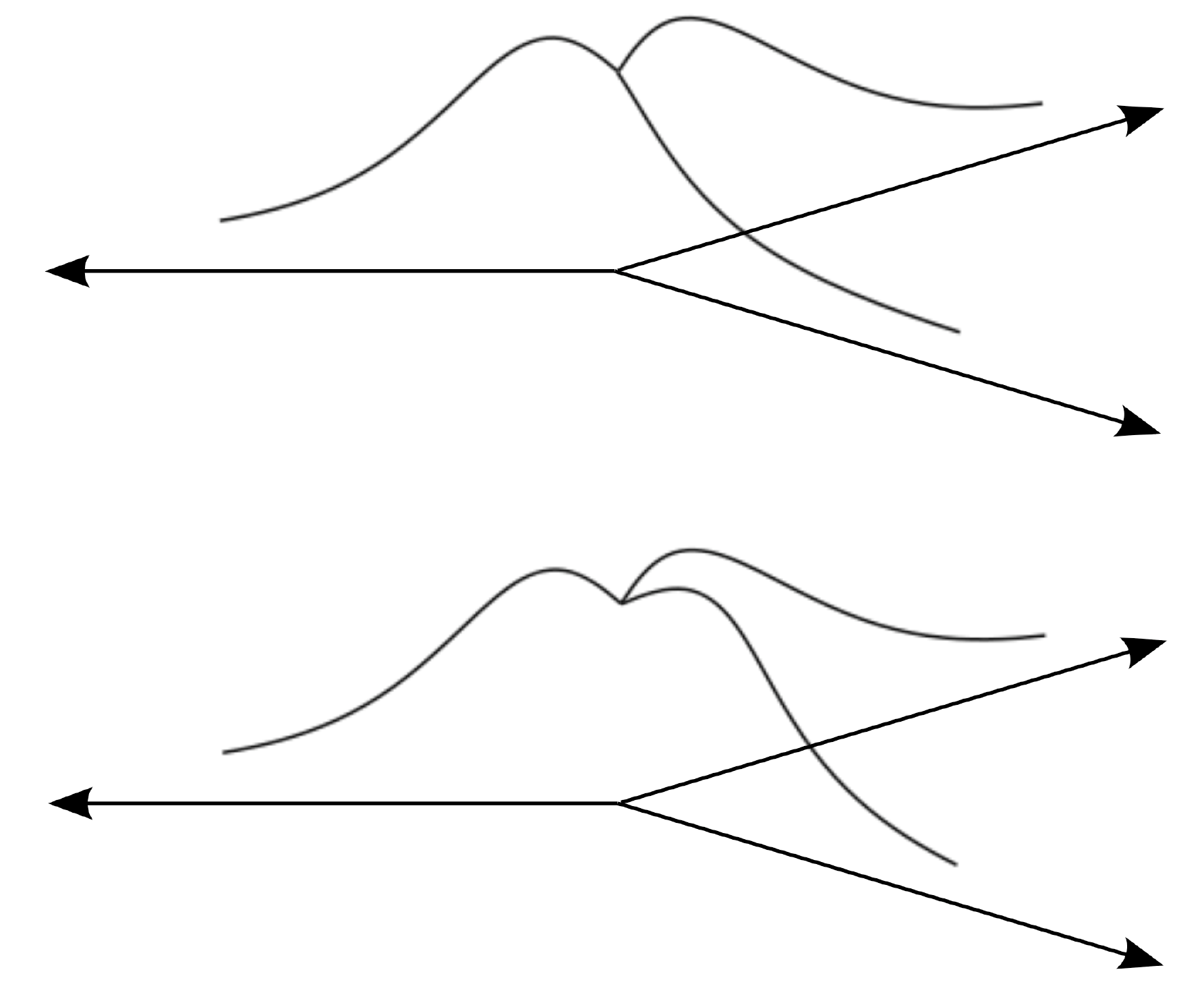}
\caption{\label{fig2}Nonlinear stationary states:  $\alpha>0, N=3, j=2,3$}
\end{center}
\end{figurehere}

Let us now consider the free case, $\al=0$. If $N$ is odd the only solution to \eqref{eq-a} is $a=0$, and the stationary state $\hat\Psi_{\omega}$ is
unique and it is given by 
\beq 
\left(\hat\Psi_{\omega}\right)_i(x) = \phi(0,x),
\quad i=1,\ldots,N.  
\eeq This corresponds to $N$ half solitons glued at the vertex.
If $N$ is even a family of solutions is given by
\beq
\lf(\hat\Psi_{\omega}^a\ri)_i(x) = 
\begin{cases}
\phi(-a,x),\quad & i=1,\ldots N/2 \\
\phi(+a,x), & i=N/2+1, \ldots N
\end{cases} \quad a\in\erre
\eeq 
Here $a$ is a free parameter. In this case the graph can be
considered just as a set of $N/2$ copies of the real line. The above solutions
$\hat\Psi_{\omega}^a$ can be interpreted as $N/2$ identical NLS
solitary waves on each real line translated by a quantity $a$. We remark finally
that from this last family of stationary states one can construct,
through Galilean invariance and on each copy of the line, traveling
waves of the form \beq \Psi_{tr} (t)= e^{i(\frac{v}{2} x-\frac{v^2}{4}
  t+\theta)} \hat\Psi_{\omega}^{a(t)} \qquad a(t)= a+vt\,. \eeq 
\noindent
\section{Ground state and variational properties}
Now we want to discuss existence and identification of the ground state of the system among
the domain ${\mathcal E}$, defined as the space of square integrable functions with square integrable derivative on every edge (sometimes indicated in the mathematical literature as $H^1(\GG)$, the first Sobolev space on the graph) and continuous in the vertex. The
stationary states of the NLS on graphs previously determined are critical points ($S_{\omega}'[\Psi]=0$) of the so called {\it action functional} $S_{\omega}$
\beq
\label{action}
\begin{aligned}
S_{\omega}[\Psi]=E[\Psi]+\frac{\omega}{2} ||\Psi||^2_2 \ .
\end{aligned}
\eeq 
For a BEC the action is nothing  but the grand potential functional (at zero temperature) and the parameter $\omega$, usually denoted as $\mu$ is the chemical potential. The {\it ground state} is the stationary state on which the action is minimal. For the main example of attracting defect this is the $N$ tail state, but this does not mean that it is a minimum of the action. The action, due to the negative nonlinear part of the energy, is unbounded from below, and to study its minima it is necessary to constrain it suitably. So we minimize $S_{\omega}$ on the so called natural constraint 
$
\{\Psi \in {\mathcal E}\ {\rm s.t.}\  I_{\omega}[\Psi]\equiv \langle S_{\omega}'[\Psi],\Psi\rangle=0\}\ ,
$
which contains all the stationary states by definition, and where the action turns out to be bounded from below.
One gets the following result, the non trivial proof of which will be given elsewhere (see \cite{ACFN3}  where the needed mathematical tools to get the result are given in detail).

\begin{theorem}
\label{th1}
Let $\mu>0$; there exists $\alpha^*<0$ such that for $\alpha<\alpha^*$ and $\omega>\frac{{\alpha}^2}{N^2}$, the action $S_{\omega}$ attains its constrained  minimum on the $N$ tail state $\Psi^{0}_{\omega}$.
\end{theorem}

 The limitation in $\alpha$ is due to the fact that for $\alpha=0$ the $N$-tails state is not a minimum but a saddle of the energy functional. This is shown in \cite{ACFN2}, constructing a trial wavefunction given by a quasi soliton almost completely concentrated on a single edge with an energy lower than the energy of a stationary state. To avoid this difficulty and guarantee the minimizing character of the stationary state given in the Theorem \ref{th1} a sufficiently strong potential well is needed
 
 A second physically relevant constrained problem is to minimize the energy at constant mass, i.e.
\beq
\label{minenergy}
\inf\{E[\Psi]\ {\rm s. t.}\ \Psi\in {\mathcal E},
M[\Psi]=M[\Psi^0_{\omega}] \}\ .
\eeq  
Here we only give an explicit example for the especially
interesting case of the cubic NLS.  A direct calculation shows that
the mass of the bound states $\Psi^j_{\omega}$ is independent of $j$:
$
||\Psi^j_{\omega}||_2^2=2N{\omega}^{\frac{1}{2}} + 2\alpha\ .
$
Notice that this degeneration is a special property of cubic case, not shared in general by different nonlinearities.
This makes the energy spectrum at fixed mass exactly computable:
\beq
\label{energyspectrum}
E[\Psi^j_{\omega}]=-\frac{N}{3}{\omega}^{\frac{3}{2}} - \frac{1}{3}\frac{{\alpha}^3}{(2j-N)^2}\ .
\eeq  
The $N$ bound states (formula holds true for both signs of $\alpha$) have an energy increasing with the number of bumps.
A closer analysis shows that this ordering holds true for the general power nonlinearity, but in this case it is not possible to give an explicit nonlinear energy spectrum.

We end this section noticing that for $\alpha<0$ the only stationary state respecting the symmetry of the interaction at the vertex is the ground state, while in the excited states the symmetry with respect to the edges is broken.

\section{Stability of the ground state} Now we come to the stability of the ground state. Due to
  the $U(1)$, or phase, invariance of \eqref{diffform} the
  stability has to be considered as {\it orbital} stability \cite{GSS,SS,W2}, which means ordinary Lyapunov stability {\it up to symmetries}: a solution remains as close as desired {\it to the orbit} $\left\{e^{i\omega t}\ \Psi_{\omega}\right\}$ of a ground state if it starts close enough to it. Establishing orbital stability is generally more difficult of establishing stability of an isolated equilibrium point. An analogous study for the delta potential on the line was given in \cite{CM}, second reference. As in
  the scalar case, the NLS on a graph turns out to be a hamiltonian
  system on the real Hilbert space of the couples of real and
  imaginary part of the wavefunction.  Decomposing $\Psi=u+iv$, \eqref{diffform} is
  equivalent to the canonical system \beq
\label{hamilt}
\frac{d}{dt} 
\begin{pmatrix}
u \\ v
\end{pmatrix}={\mathcal J} E'[u,v]\ ,\qquad {\mathcal J}=
\begin{pmatrix}
\ 0\ \ I \\ -I\ 0
\end{pmatrix}.
\eeq
where the Hamiltonian $E[u,v]$ is obtained from \eqref{energy} after posing $\Psi=u+iv\ .$
Linearization of the hamiltonian system \eqref{hamilt} around the stationary state is achieved by substituting
$
(\Psi (t))_k=(\Psi^0_{\omega,k} + \eta_k +i\rho_k)e^{i\omega t}
$
and neglecting higher order terms than linear in \eqref{hamilt}. The real vector functions $\eta$ and $\rho$ satisfy
 \beq
\label{linhamsyst}
\frac{d}{dt} 
\begin{pmatrix}
\eta \\ \rho
\end{pmatrix}={\mathcal J} {\mathcal L}
\begin{pmatrix}
\eta \\ \rho
\end{pmatrix}\ .
\eeq
where ${\mathcal L}={\rm diag}({\mathcal L}_-,{\mathcal L}_+)$ and  
\beq
\label{linhamilt}
\begin{aligned}
\left({\mathcal L}_+\right)_{i,k} &=\left(-\frac{d^2}{dx^2} + \omega - |\Psi^0_{\omega,k}|^{2\mu}\right){\delta}_{i,k} \\ 
\left({\mathcal L}_-\right)_{i,k} &=\left(-\frac{d^2}{dx^2} +\omega - (2\mu+1)|\Psi^0_{\omega,k}|^{2\mu}\right){\delta}_{i,k} \ .
\end{aligned}
\eeq ${\mathcal L}_-$ and ${\mathcal L}_+$ are matrix self adjoint
operators acting on the real vector functions $\eta$ and $\rho$ satisfying \eqref{domdelta}.  Note that ${\mathcal L}_+{\Psi_{\omega}^0}=0$. Precise conditions to
have orbital stability for general hamiltonian
systems and in particular for NLS equations are now 
classical \cite{W2, GSS}. They are based on the properties
of linearization $\mathcal L$ and on the so called {\it
  Vakhitov-Kolokolov} (\cite{VK}) criterion. These results imply in particular
that solitary solutions of \eqref{hamilt} are orbitally stable if\ i)
{\it spectral} conditions hold: $i_1$) $\ker {\mathcal L}_+= \{\Psi^0_{\omega}\}$ and
the rest of the spectrum is positive; $i_2$) $n({\mathcal L}_-)=1$
where the left hand side is the number of negative eigenvalues.  ii)
{\it Vakhitov-Kolokolov} condition $\frac{d}{d\omega}||\Psi^0_{\omega}||^2_2 > 0$
holds. 

\begin{theorem} 
Let $\mu \in [0,2]$,
$\alpha<\alpha^*<0$, $\omega>\frac{{\alpha}^2}{N^2}$. Then the ground state
$\Psi^{0}_{\omega}\ $ is orbitally stable in ${\mathcal E}$.
\end{theorem}

Concerning Vakhitov-Kolokolov condition one has 
\beq
\begin{aligned}
\label{VK}
&\frac{d}{d\omega}||\Psi^0_{\omega}||^2_2=
C\bigg[\left(\frac{1}{\mu}-\frac{1}{2}\right)\int^1_{\frac{|\alpha|}{N\sqrt{\omega}}} (1-t^2)^{\frac{1}{\mu}-1}\ \upd t \\&+\frac{|\alpha|}
{2N\sqrt{\omega}}\left(1-\frac{|\alpha|^2}{N^2\omega}\right)^{\frac{1}{\mu}-1}\bigg] >0
\end{aligned}
\eeq with
$C=C(N,\mu,\omega)=N\frac{(\mu+1)^{\frac{1}{\mu}}}{\mu}\omega^{\frac{1}{\mu}-\frac{3}{2}}$.
  Concerning $i_1$), $\Psi^0_{\omega}$ is positive, and we prove now that it is
the (simple) ground state of ${\mathcal L}_+$.  
An integration by parts allows to rewrite 
\beq
\langle {\mathcal L}_+ \Psi, \Psi\rangle= \sum_{k=1}^N\int_0^{+\infty} ({\Psi^0}_{\omega,k})^2 |\frac{d}{dx}(\frac{\psi_k}{\Psi^0_{\omega,k}})|^2 \upd x
\eeq
where the finite term in the integration is vanishing due to the $\delta$ boundary condition.
So $\langle {\mathcal L}_+ \Psi,
\Psi\rangle > 0 $ for every $\Psi$ not coinciding with $\Psi^0_{\omega}$, which is the only eigenvector with eigenvalue $0$. Concerning
$i_2$), note that the ground state is a
strict local minimum of $S_{\omega}$ constrained on the codimension one 
natural manifold $I^{-1}_{\omega}(0)$. So ${\mathcal L}$ is
positive on the tangent space to $I^{-1}_{\omega}(0)$ at the ground
state.
Correspondingly, ${\mathcal L}$ could have at most one negative
eigenvalue.  In fact it has one, being $\langle{\mathcal L}_-\Psi^0_{\omega},\Psi^0_{\omega}\rangle<0$.  This ends the proof.
\par\noindent 
We add some remarks. In the first place notice that from formula \eqref{VK} it follows that for $\mu>2$ there exists $\omega^*$ such that $\Psi^{0}_{\omega}\ $ is orbitally stable for $\omega\in (\frac{{\alpha}^2}{N^2},\omega^*) $ and is orbitally unstable for $\omega>\omega^*\ .$ So one has orbital stability of the branch of ground states also in the supercritical regime $\mu>2$, where global existence of solutions is not guaranteed and the solution could blow up in finite time, if the frequency $\omega$ of the standing wave is not too high. \par\noindent Nothing rigorous is known about orbital stability or instability of excited states; it is expected that excited states are unstable.

\section{Conclusions}
We begin with this letter the analysis of nonlinear wave propagation
on ramified structures and networks. We have considered a star graph
($N$ halflines, ``edges'', attached at a single vertex) which sustains a
nonlinear focusing Schr\"odinger dynamics of power type. Interaction
between the fields on different edges is through a boundary condition
at the vertex. For this system explicit nonlinear bound states exist when
certain classes of boundary conditions ($\delta$ type, attractive or
repulsive) are considered. In the most common case of Kirchhoff
vertex, for an even number of edges only, traveling waves exist.  
Let
us stress the fact that there exist excited states beside the ground
state,  
For example they give rise
to a nonlinear energy spectrum for the system, here displayed in the
case of cubic nonlinearity.  Moreover, the ground states have a
variational characterization and they are orbitally stable for
sufficiently attractive vertex and nonlinearity below a critical
value. The study can be extended to the NLS with different
nonlinearities, for example to combination of several power
nonlinearities, such as the important case of cubic-quintic
case. Moreover similar techniques can be applied to other nonlinear
dispersive equations admitting solitons solutions or standing waves.
Finally, apart from the possible applications of the model to the
description of real systems involving propagation on networks, the
obtained results rise several general interesting issues.  We mention
two of them, which seem to be important for a general comprehension of
the dynamics of solitary waves. The first is the analysis of stability
and instability of excited states of the system. It is known that
excited standing waves exist in many nonlinear models, but it is in
general quite difficult to study them due to the rather implicit
information at disposal. The present model has explicit excited
states, and it appears as promising their stability study, in the
first place at a linearized level (given by Eq. \eqref{linhamsyst}) , and then concerning orbital
(in)stability.  A second interesting issue concerns the asymptotic
stability of solitons, i.e. the fact (to be proven in every single
model) that the evolution of an initial state in a suitable class
asymptotically decomposes in one or more solitary waves, radiation
driven by the linear evolution and a small remainder. When true this
property is obviously quite important from the physical point of view
because it yields to a strong  simplification of the behaviour of the
system in the asymptotic regime. A possible proof is related in the
first place to the spectral properties of linearization ${\mathcal J}
{\mathcal L}$ and to the decay properties of solutions of the linearized equation
\eqref{linhamsyst}, and again one can hope to exploit the rather
explicit expressions of these objects.

\thanks{The authors would like to thank G.N.F.M. (Gruppo Nazionale per la Fisica Matematica) for financial support and R. Fukuizumi for discussions. R.A. is partially supported by the PRIN2009 grant  ``Critical Point Theory and Perturbative Methods for Nonlinear Differential Equations''}

\end{multicols}
  

\begin{thebibliography}{10}
\bibitem{SS} Sulem C., Sulem P., {\it The nonlinear Schr\"odinger equation} (Springer, New York) 1999.
\bibitem{CM} Cao X.D., Malomed B.A., {\it Phys. Lett. A}, {\bf 206} (1995) 177. 
\bibitem{F} Le Coz S. {\it et al}, {\it Physica D}, {\bf 237} (2008) 1103. 
\bibitem{Linzon} Linzon Y. {\it et al.}, {\it Phys. Rev. Lett.}, {\bf 99} (2007) 133901. 
\bibitem{Pe} Peccianti M.  {\it et al.}, {\it Phys. Rev. Lett.}, {\bf 101} (2008) 153902.  
\bibitem{LB} Lee C., Brand J., {\it Europhys. Lett.}, {\bf 73} (2006) 321. 
\bibitem{Wan} Wan, W. {\it et al.},  {\it Phys. Rev. Lett.},  {\bf 104} (2010) 073903. 
\bibitem{TOD} Tokuno A. {\it et al.}, {\it Phys. Rev. Lett.},  {\bf 100} (2008) 140402. 
\bibitem{BS} Bulgakov E., Sadrev A., {\it Phys. Rev. E}, {\bf 84} (2011) 155304. 
\bibitem{Miro} Miroshnichenko A.E. {\it et al.},  {\it Phys. Rev. E}, {\bf 75} (2007) 046602.  
\bibitem{S} Sobirov Z. {\it et al.}, {\it Phys. Rev. E}, {\bf 81} (2010) 066602. 
\bibitem{GSD} Gnutzmann S., Smilansky U., Derevyanko S., {\it Phys. Rev. A}, {\bf 83} (2011) 033831. 
\bibitem{ACFN} Adami R., Cacciapuoti C., Finco D., Noja D.,  {\it Rev. Math. Phys}, {\bf 23} (2011) 4.  
\bibitem{EKK} Exner P. {\it et al.} (Editor), \emph{Analysis on graphs and its applications},  Proceedings of Symposia in Pure Mathematics, Vol. {\bf 77} (AMS) 2008.
\bibitem{B} Berg\'e L., {\it Phys. Rep.}, {\bf 303} (1998) 259.
\bibitem{ACFN3} Adami R., Cacciapuoti C., Finco D., Noja D., arXiv:1206.5201 (2012).
\bibitem{ACFN2} Adami R., Cacciapuoti C., Finco D., Noja D., {\it J. Phys. A: Math. Theor.}, {\bf 45} (2012) 192001.
\bibitem{W2} Weinstein M., {\it Comm. Pure Appl. Math.}, {\bf 39} (1986) 51.  
\bibitem{GSS} Grillakis M., Shatah J., Strauss W., {\it J. Funct. Anal.}, {\bf 74} (1987) 160. 
\bibitem{VK} Vakhitov M.G., Kolokolov A.A., {\it Radiophys. Quantum Electron},  {\bf 16} (1973) 783. 
\end{thebibliography}
\end{document}